\documentclass[]{emulateapj}

\usepackage{subfigure}
\usepackage{url}
\usepackage{hyperref}
\usepackage{longtable}
\usepackage{natbib}
\usepackage{amsmath}
\usepackage[normalem]{ulem}
\usepackage{bm}
\usepackage{array}
\newcolumntype{P}[1]{>{\centering\arraybackslash}p{#1}}
\newcolumntype{M}[1]{>{\centering\arraybackslash}m{#1}}
\usepackage[usenames, dvipsnames]{color}

\newcommand{\noprint}[1]{}

\usepackage{color}

\newcommand{\kep}{{\it Kepler}}

\newcommand{\mearth}{{M$_\oplus$}}

\newcommand{\rsun}{{R$_\odot$}}
\newcommand{\msun}{{M$_\odot$}}
\newcommand{\mjup}{{M$_\textrm{Jup}$}}

\newcommand{\thisstar}{Kepler-56}

\newcommand{\mstar}{{$M_\star$}}

\begin{document}
\title{The Orbit and Mass of the Third Planet in the Kepler-56 System}

\newcommand{\princeton}{1}
\newcommand{\caltech}{2}
\newcommand{\cfa}{3}
\newcommand{\standrews}{4}
\newcommand{\berkeley}{5}
\newcommand{\geneva}{6}
\newcommand{\palermo}{7}
\newcommand{\galileob}{8}
\newcommand{\milano}{9}
\newcommand{\galileo}{10}
\newcommand{\padova}{11}
\newcommand{\cavendish}{12}
\newcommand{\edinburgh}{13}
\newcommand{\torino}{14}
\newcommand{\queens}{15}

\author{%
Oderah~Justin~Otor\altaffilmark{\princeton}, 
Benjamin~T.~Montet\altaffilmark{\caltech,\cfa}, 
John~Asher~Johnson\altaffilmark{\cfa},
%HARPS-N & HIRES Teams, Alphabetized
David~Charbonneau\altaffilmark{\cfa},
Andrew~Collier-Cameron\altaffilmark{\standrews},
Andrew~W.~Howard\altaffilmark{\caltech},
Howard~Isaacson\altaffilmark{\berkeley},
David~W.~Latham\altaffilmark{\cfa},
Mercedes~Lopez-Morales\altaffilmark{\cfa},
Christophe~Lovis\altaffilmark{\geneva},
Michel~Mayor\altaffilmark{\geneva},
Giusi~Micela\altaffilmark{\palermo},
Emilio Molinari\altaffilmark{\galileob,\milano},
Francesco~Pepe\altaffilmark{\geneva},
Giampaolo~Piotto\altaffilmark{\galileo,\padova},
David~F.~Phillips\altaffilmark{\cfa},
Didier~Queloz\altaffilmark{\geneva,\cavendish},
Ken~Rice\altaffilmark{\edinburgh},
Dimitar~Sasselov\altaffilmark{\cfa},
Damien~S\'egransan\altaffilmark{\geneva},
Alessandro~Sozzetti\altaffilmark{\torino},
St\'ephane~Udry\altaffilmark{\geneva},
Chris~Watson\altaffilmark{\queens}
}

\email{ootor@princeton.edu}

\altaffiltext{\princeton}{Department of Astrophysical Sciences, Princeton University, 4 Ivy Lane, Princeton, NJ 08544, USA}
\altaffiltext{\caltech}{Cahill Center for Astronomy and Astrophysics, California Institute of Technology, 1200 E. California Blvd., MC 249-17, Pasadena, CA 91106, USA}
\altaffiltext{\cfa}{Harvard-Smithsonian Center for Astrophysics, 60 Garden Street, Cambridge, MA 02138, USA}
%\altaffiltext{\ifa}{Institute for Astronomy, University of Hawai'i at M\={a}noa, Hilo, HI 96720, USA}
\altaffiltext{\standrews}{SUPA, School of Physics \& Astronomy, University of St. Andrews, North Haugh, St. Andrews Fife,
KY16 9SS, UK}
\altaffiltext{\berkeley}{Department of Astronomy, University of California, Berkeley CA 94720, USA}
\altaffiltext{\geneva}{Observatoire Astronomique de l'Universit\'e de Gen\`eve, 51 ch. des Maillettes, 1290 Versoix, Switzerland}
\altaffiltext{\palermo}{INAF - Osservatorio Astronomico di Palermo, Piazza del Parlamento 1, 90134 Palermo, Italy}
\altaffiltext{\galileob}{INAF - Fundacion Galileo Galilei, Rambla Jose Ana Fernandez Perez 7, 38712 Brena Baja, Spain}
\altaffiltext{\milano}{INAF - IASF Milano, via Bassini 15, 20133, Milano, Italy}
\altaffiltext{\galileo}{Dipartimento di Fisica e Astronomia ``Galileo Galilei'', Universita'di Padova, Vicolo dell'Osservatorio 3, 35122 Padova, Italy}
\altaffiltext{\padova}{INAF - Osservatorio Astronomico di Padova, Vicolo dell'Osservatorio 5, 35122 Padova, Italy}
\altaffiltext{\cavendish}{Cavendish Laboratory, J. J. Thomson Avenue, Cambridge CB3 0HE, UK}
\altaffiltext{\edinburgh}{SUPA, Institute for Astronomy, University of Edinburgh, Royal Observatory, Blackford Hill, Edinburgh,
EH93HJ, UK}
\altaffiltext{\torino}{INAF - Osservatorio Astrofisico di Torino, via Osservatorio 20, 10025 Pino Torinese, Italy}
\altaffiltext{\queens}{Astrophysics Research Centre, Queen's University Belfast, Belfast BT7 1NN, UK}

%\date{\today, \currenttime}

\begin{abstract}
 While the vast majority of multiple-planet systems have orbital angular momentum axes that align with the spin axis of their host star, Kepler-56 is an exception: its two transiting planets are coplanar yet misaligned by at least 40 degrees with respect to the rotation axis of their host star. Additional follow-up observations of Kepler-56 suggest the presence of a massive, non-transiting companion that may help explain this misalignment. We model the transit data along with Keck/HIRES and HARPS-N radial velocity data to update the masses of the two transiting planets and infer the physical properties of the third, non-transiting planet. We employ a Markov Chain Monte Carlo sampler to calculate the best-fitting orbital parameters and their uncertainties for each planet. We find the outer planet has a period of $1002 \pm 5$ days and minimum mass of $5.61 \pm 0.38$ \mjup. We also place a 95\% upper limit of 0.80 m s$^{-1}$ yr$^{-1}$ on long-term trends caused by additional, more distant companions.
\end{abstract}


\maketitle
\section{Introduction}
\label{sec:intro}

Red giant \thisstar\ (KOI-1241, KIC\,6448890) is an atypical star to host transiting planets.
While the vast majority of known transiting planets orbit solar-type FGK stars
\citep{Batalha13, Burke14, Mullaly15, Rowe15, Grunblatt16, VanEylen16}, \thisstar\ is one of only a few post-main sequence stars known to host them \citep{Lillo-Box14, Ciceri15, Quinn15, Pepper16}. Detecting transits of these stars is difficult because they are much larger than main sequence stars and have higher levels of correlated noise \citep{Barclay15}. As such, when selecting targets for \kep\,, mission scientists prioritized capturing main sequence FGK stars over other stellar types \citep{Batalha10}.

Nevertheless, \thisstar\ was targeted in the original \kep\ mission \citep{Borucki10}, and two transiting planet candidates with periods of $10.50$ and $21.41$ days were identified in the
first data release \citep{Borucki11b}. These candidates interacted dynamically, with observed, anticorrelated variations in their times of transit \citep{Ford11, Ford12TTV, Steffen12}.
\citet{Steffen13} analyzed the times of transit and the orbital stability of the system to confirm these two candidates as planets, making
Kepler-56 the latest stage star known at the time to host multiple
transiting planets.

As a red giant, \thisstar\ exhibits convection-driven oscillations that vary on timescales long enough to be observable with \kep\ long-cadence photometry.
\citet{Huber13} analyzed its observed asteroseismic modes
to infer a stellar mass of $1.32 \pm 0.13$ \msun\ and radius of $4.23 \pm 0.15 $ 
\rsun.
Through radial velocity (RV) and transit timing observations of the transiting planets, \citet{Huber13} then determined their masses to be $22.1^{+3.9}_{-3.6}$ \mearth\ and 
$181^{+21}_{-19}$ \mearth, respectively.
Through a combination of asteroseismology and dynamical instability simulations, they also detected that the orbits of the planets, while coplanar with each other, are tilted with respect to the axis of stellar rotation by $\sim 45$ degrees.

\citet{Huber13}\ also detected the presence of a long-term RV acceleration in the data consistent with at least one additional massive companion. While the acceleration by itself
cannot provide a unique orbit for the outer companion, they proposed that both the planetary obliquity and long-term RV trend could be explained by a non-transiting companion with a period of $900$ days and mass $3.3$ \mjup.

However, the duration of their RV observations only covered a baseline of $\sim 100$ days. Equipped with four more years of RV data, we are now able to measure the orbital parameters of this purported planet, which has the third-longest orbital period of any confirmed planet orbiting a Kepler star \citep{Kostov16, Kipping16}. We are also able to place upper limits on the presence of additional planets from the lack of additional long-term trends in the RV curve.

In Section \ref{sec:data} we describe our data collection and reduction. 
In Section \ref{methods}, we describe our RV model. In Section \ref{results}, we present our best estimates for this planet's orbital parameters, as well as the likelihood of another companion. We discuss our results in Section \ref{discussion} and summarize our findings in Section \ref{summary}.

\section{Data Collection and Analysis}
\label{sec:data}

Our analysis is based on 43 RV observations of Kepler-56 obtained from 2013 to 2016 with two different spectrographs: 24 with Keck/HIRES \citep{Vogt94} and 19 with HARPS-North \citep{Cosentino12}.

\subsection{Keck/HIRES Observations}

Our Keck/HIRES observations were obtained largely following the standard procedures of the California Planet Survey (CPS) team \citep{Howard10}, modified slightly for the faint stars of the \kep\ field, following the approach of 
\citet{Huber13}.
For all observations, we used the C2 decker (14\farcs0 $\times$ 0\farcs85), which is a factor of four taller than the B5 decker typically used for 
observations of bright stars. 
This setup allows for more background light to enter the spectrograph, allowing for better sky-subtraction while maintaining a resolving power of $R \approx $ 50,000. 

Each observation was made with an iodine cell mounted along the light path before the entrance to the spectrograph. The iodine spectrum superposed on the stellar spectrum provides a precise, stable wavelength scale and information on the shape of the instrumental profile of each observation \citep{Valenti95,Butler_etal96}.

The integration times range from 600 to 1800 seconds. The star-times-iodine spectrum was modeled using the \citet{Butler_etal96} method, with the instrumental profile removed through numerical deconvolution. The RV of the star at each observation is compared to a template spectrum
of the star obtained without iodine, with the instrumental profile removed through numerical deconvolution. The observed RVs are listed in
Table \ref{tab:data}.

The data set used here includes the 10 observations used by \citet{Huber13}, reanalyzed after all observations were recorded.
An improved stellar template spectrum causes the measured RV from these observations to be slightly different than those reported by \citet{Huber13}, although the differences are smaller than the formal uncertainties
on each observation.

\subsection{HARPS-North Observations}

We also obtained 19 observations of \thisstar\ with HARPS-North, a
high-precision echelle spectrograph at the 3.6 m Telescopio
Nazionale Galileo (TNG) at the Roque de los Muchachos
Observatory, La Palma, Spain.
HARPS-N is a fiber-fed high-resolution ($R=115,000$) spectrograph optimized for measuring precise RVs. 

The exposure times for all observations with HARPS-N were 1800 seconds,
and the data were reduced with version 3.7 of the standard HARPS-N pipeline.
RVs were derived with the standard weighted cross-correlation function
method \citep{Baranne96, Pepe02}. 
These data are also listed in Table \ref{tab:data}. 

Note that the HARPS-N pipeline includes the systemic RV, while the Keck/HIRES pipeline does not, leading to a 54.25 km s$^{-1}$ apparent shift between the two sets.

\begin{table}[h!]
    \centering
    \caption{RV Observations of Kepler-56}
    \label{tab:data}
    \begin{tabular}{P{2.3 cm} P{1.6 cm} P{1.6 cm} P{1.6 cm}}
        \hline
        Time (BJD-2,400,000) & RV (m $s^{-1}$) & RV uncertainty & Spectrograph \\
        \hline
        56076.904 & -38.30 & 2.51 & HIRES \\
        56099.841 & -13.18 & 2.47 & HIRES \\
        56109.825 & 57.33 & 1.74 & HIRES \\
        56116.089 & -4.45 & 1.56 & HIRES \\
        56134.000 & 46.27 & 1.73 & HIRES \\
        56144.079 & 17.81 & 2.02 & HIRES \\
        56153.087 & 88.74 & 3.29 & HIRES \\
        56163.981 & 37.40 & 1.91 & HIRES \\
        56166.962 & 44.33 & 1.83 & HIRES \\
        56176.856 & 88.23 & 2.29 & HIRES \\
        56192.844 & 108.96 & 1.86 & HIRES \\
        56450.040 & 19.85 & 1.78 & HIRES \\
        56469.099 & 3.95 & 1.86 & HIRES \\
        56472.114 & 16.87 & 1.99 & HIRES \\
        56476.995 & 1.60 & 2.03 & HIRES \\
        56478.884 & -29.12 & 1.65 & HIRES \\
        56484.063 & -72.43 & 2.00 & HIRES \\
        56484.883 & -72.22 & 1.50 & HIRES \\
        56489.997 & -13.62 & 1.62 & HIRES \\
        56506.878 & -84.42 & 1.78 & HIRES \\
        56512.910 & -14.65 & 1.77 & HIRES \\
        56521.883 & -54.61 & 1.73 & HIRES \\
        56533.873 & -38.28 & 2.05 & HIRES \\
        56613.758 & -99.60 & 2.23 & HIRES \\
        \hline
        56462.573 & -54305.87 & 4.45 & HARPS-N \\
        56514.602 & -54269.66 & 2.92 & HARPS-N \\
        56514.623 & -54258.05 & 3.05 & HARPS-N \\
        56515.556 & -54259.96 & 4.82 & HARPS-N \\
        56515.557 & -54271.49 & 5.02 & HARPS-N \\
        56515.578 & -54258.19 & 4.35 & HARPS-N \\
        56545.423 & -54331.28 & 2.42 & HARPS-N \\
        56549.407 & -54343.28 & 3.12 & HARPS-N \\
        56829.617 & -54375.25 & 3.16 & HARPS-N \\
        56831.525 & -54359.40 & 2.30 & HARPS-N \\
        56850.615 & -54386.77 & 4.12 & HARPS-N \\
        56865.533 & -54359.15 & 2.21 & HARPS-N \\
        57123.719 & -54256.32 & 4.55 & HARPS-N \\
        57181.709 & -54148.86 & 2.71 & HARPS-N \\
        57254.564 & -54202.52 & 6.42 & HARPS-N \\
        57330.394 & -54147.12 & 3.19 & HARPS-N \\
        57528.706 & -54303.21 & 4.78 & HARPS-N \\
        57565.651 & -54280.83 & 5.24 & HARPS-N \\
        57566.674 & -54290.92 & 3.12 & HARPS-N \\

    \end{tabular}
\tablecomments{The Keck/HIRES pipeline returns RVs with the systemic
RV, $\gamma$, removed; this offset is retained in the HARPS-N RVs, leading to an apparent shift of 54.25 km s$^{-1}$.}
\end{table}

\section{Orbit Fitting}
\label{methods}

With the RV data in hand, we can determine the orbital parameters of the outer planet. We develop code that, for a given set of orbital parameters, returns the expected RV contribution from each planet at a list of user-specified times following \citet{LehmannFilhes94} and \citet{Eastman13}.

Our algorithm does not include variations caused by dynamically interacting planets. However, \thisstar\,b's RV signal is small relative
to our RV precision and the magnitude of \thisstar\,c's perturbation
is small relative to its orbital period, so we do not expect to see any
perturbation signal in the data.
The two spectrograph pipelines return different RV offsets, so we make an initial guess for the relative offset between the two in our fitting. 

For each planet, we include the minimum mass ($m \sin{i}$), including the unknown inclination of the non-transiting planet, and two vectors which define the eccentricity and argument of periastron ($\sqrt{e}\cos{\omega}$ and $\sqrt{e}\sin{\omega}$), following \citet{Eastman13}.

For the outer planet alone, we include orbital period ($P$) and time of transit ($t_{\textrm{tr}}$, if it were so aligned); these values are fixed for the inner planets. Stellar mass (\mstar), separate instrumental offsets ($\gamma$), and RV jitter terms ($\sigma_{jitter}$) for HARPS and HIRES complete our list of parameters. 
Functionally, as the HARPS pipeline returns a measurement with the systemic RV included (ignoring features like the gravitational redshift and convective blueshift), the offset associated with that instrument approximates the true systemic velocity of the star while the offset for HIRES brings these two sets of observations onto the same scale.

We only consider models of three planets plus a long-term RV acceleration.
While it is possible that two planets in circular orbits with orbital periods near a 2:1 period ratio can
masquerade in RV observations as a single planet with a higher eccentricity \citep{Anglada-Escude10}, 
there is no evidence that such an effect is occurring in our data set.
However, we lack the phase coverage to fully rule out this hypothesis.
More observations where our coverage is sparse would be helpful to 
probe for a fourth planet in resonance with the third.

After solving Kepler's equation to obtain the Keplerian orbital elements, the function produces radial velocities following:
\begin{equation}
\label{eq:rv}
\begin{split}
\textrm{RV} &= \bigg(\frac{2 \pi G}{P}\bigg)^{1/3} \frac{m \sin{i}}{(M_\star+m)^{2/3}}\frac{1}{\sqrt{1 - e^2}} \\
 &\times \bigg[\cos{(\theta(e,\omega,t_{\textrm{obs}},t_{\textrm{obs}}) + \omega)} + e \cos \omega \bigg].
 \end{split}
\end{equation}
Here, $\theta$ represents the true anomaly, $t_{\textrm{obs}}$ is its specific value, and $\omega$ is the argument of periastron.

With our function's ability to generate an RV curve for any specified period, we can test various combinations of the outer companion's orbital parameters. We exploit this ability in performing successive fits to obtain an initial estimate of our planetary parameters.

\subsection{Maximum Likelihood Estimation}
\label{mle}

First, we perform maximum likelihood estimation via Python's \texttt{scipy.optimize.minimize} routine. For the possible companion, we take all values as unknown. Specifically, we fit for $\sqrt{e} \cos \omega$, $\sqrt{e} \sin \omega$, $m \sin i$, $P$, \mstar, $t_\textrm{tr}$, and $\dot{\gamma}$, the acceleration of the entire system over time. Since our measurements come from two instruments, we include independent offset terms for each, $\gamma_{HARPS}$ and $\gamma_{HIRES}$, where $\gamma$ is the systemic RV offset term introduced in Section \ref{methods}. There are 17 free parameters in total -- these, plus $\sqrt{e}\cos{\omega}$, $\sqrt{e}\cos{\omega}$, and $m \sin{i}$ for each planet (as mentioned in Section \ref{methods}).

Maximum likelihood estimation is a process in which we calculate the logarithm of likelihood ($L$) by comparing our data ($D$) to the sum of our generated RV curves through the standard equation:

\begin{equation}
\label{eq:logL} \ln{(L)} = -\frac{1}{2} \sum_{i}^{N-1} \bigg(\frac{D_i - \textrm{RV}_i}{\sigma_{jitter,i}}\bigg)^2 + \frac{1}{2} N \ln{(2 \pi \sigma_{jitter,i}^2)}\\
\end{equation}

\begin{equation}
\label{eq:s} \sigma_{jitter,i}^2 = \sigma_i^2 + j^2
\end{equation}

We use $\sigma_{jitter}$ in order to incorporate jitter. Sources of jitter include uncertainties in measurements beyond photon noise that arise from sources like noise in the detector or stellar activity. For sub-giant stars, typical jitter values are 3-5 m s$^{-1}$ \citep{Johnson08}.
Given the longer exposures for this star relative to previous studies 
of planets around relatively bright subgiants, we might expect a lower
level of jitter as the integrations will average over the higher-order
modes.

We initialize the fit with values from \citet{Huber13}.
However, we note a typo in Table 1 of the discovery paper: the listed times of
transit in that paper are too large by 20 days. They should be 
2454958.2556 and 2454958.6560 days for \thisstar\,b and c, respectively, 
rather than 2454978.2556 and 2454978.6560 days.

We reject trials with nonphysical results such as negative masses and periods. 
For steps that are not rejected, we apply normal priors with expected values and $1\sigma$ uncertainties based on measurements from \citet{Huber13} for the asteroseismic mass of the host star and the inner planets' photodynamical eccentricity vectors, based on the TTV analysis of the \kep\ light curve. The sum of the logarithm of each prior term is saved for each set of parameters that is tested.

Then, we calculate the logarithm of the posterior probability for each model, which is the sum of the log-prior and log-likelihood terms (as maximizing the logarithm of a function is equivalent to maximizing the function itself). Equation \ref{eq:postPDF} illustrates this process:

\begin{equation}
\label{eq:postPDF} 
\ln{\big[p(\textrm{RV}|D)\big]} = \ln{\big[\prod_{N} (\boldsymbol{\theta})\big]} + \ln{\big[L(D|\textrm{RV})\big]}
\end{equation}

Equation \ref{eq:postPDF} calculates the logarithm of the posterior probability distribution function for any set of model parameters ($\boldsymbol{\theta}$) as compared to our RV data ($D$). The combination of parameters found by this process to make the data most probable then becomes the initial guess for our final fitting process.

\subsection{Markov Chain Monte Carlo Analysis}

We use the result of maximum-likelihood estimation from Section \ref{mle} as the initialization for \texttt{emcee} \citep{Foreman-Mackey13}, a Markov Chain Monte Carlo (MCMC) implementation for Python of the affine-invariant ensemble sampler of \citet{Goodman10}. Our 17 parameter simulation uses 150 walkers and 6000 steps, with an observed burn-in of 1500 steps.

%\textbf{Steps to higher posterior probability regions are always taken; others} are only accepted a fraction of the time, depending on the relative posterior probabilities of the current and proposed walker positions. In this sense, the process is sensitive to its initialization. After a period called the ``burn-in," the walkers settle around the peak and the variance of predictions for each parameter consequently stabilizes. \textbf{In simpler MCMC algorithms (e.g., Metropolis-Hastings), there is only one walker, the size of its steps is not fixed, and their direction is purely random, which often increases the computation time needed to fully explore a probability distribution and decreases the efficiency at which it is explored.}

\section{Results}
\label{results}

\begin{figure*}[htb]
\includegraphics[width=\textwidth]{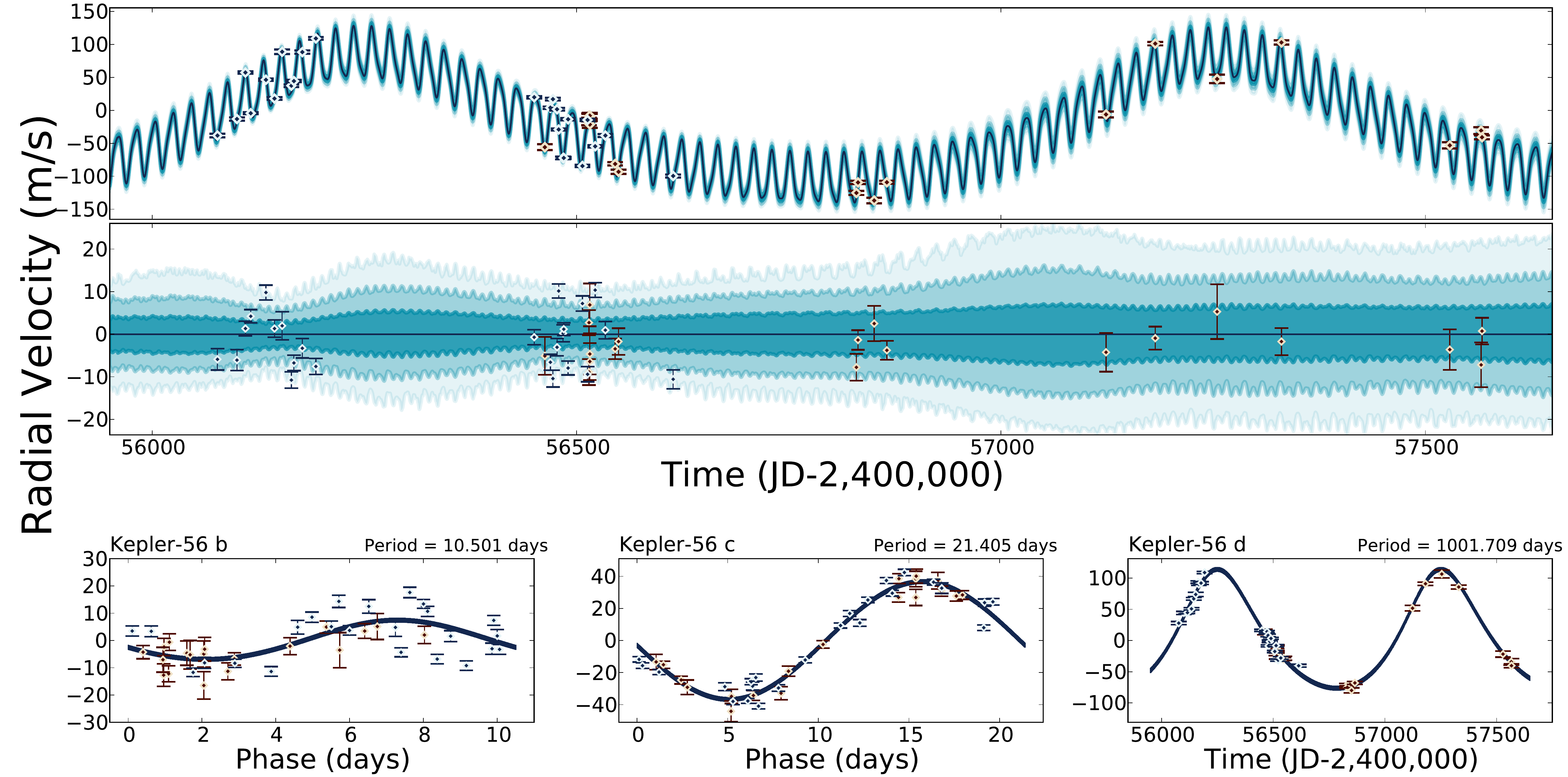}
\caption{A suite of results from our MCMC fit. (Top) \thisstar\, RV data -- with HIRES points in red and those of HARPS in navy -- together with a curve whose gradient represents the differing confidence levels of MCMC's results, with the darkest navy representing the median fit and lighter shades corresponding to the 1, 2, and 3$\sigma$ uncertainties on the RV of the star as a function of time. (Middle) Identical to Panel 1, save that each data point and confidence curve has been subtracted by the median fit RV in order to show the data residuals and uncertainty as a function of time. (Lower left) Individual, phase folded RV contribution of \thisstar\,b to the system's total RV. The contributions of the other two planets are subtracted from the HARPS and HIRES data displayed on the plot for visualization purposes. (Lower middle) The phase folded version of the lower left plot for \thisstar\,c with the signals from the other two planets removed for visualization purposes. (Lower right) \thisstar\,d's individual, non-phase folded RV contribution, again with the other two planets removed.}
\label{fig:rvs}
\end{figure*}

We detect a massive, non-transiting companion, designated \thisstar\,d, with final best-fit values and uncertainties listed in Table \ref{tab:results}. The RV curve generated by our highest-confidence combination of parameters can be seen in tandem with its uncertainties and our original RV data in Figure \ref{fig:rvs}.
In the same figure, we also show the maximum likelihood orbits for each individual planet, as well as the data
with the maximum likelihood signals from the other two planets removed.
These data are only for visualization purposes; at all times we fit the contributions 
from all three planets simultaneously.

\begin{table}[h!]
    \centering
    \caption{Orbital Parameters for the Kepler-56 System}
    \label{tab:results}
    \begin{tabular}{P{3 cm} P{1.4 cm}  P{2.3 cm}}
        \hline
        Parameters & Maximum-likelihood best-fits & \texttt{emcee} median fits \& $1\sigma$ uncertainties \\
        \hline
        \thisstar\,b & & \\
        $\sqrt{e_1}\cos{\omega_1}$ & 0.20 & 0.19 $\pm$ 0.04 \\
        $\sqrt{e_1}\sin{\omega_1}$ & -0.04 & -0.04 $\pm$ 0.05 \\
        
        $e_1$\tablenotemark{a} & 0.04 & 0.04 $\pm$ 0.01 \\
        
        $\omega_1$ (Radians)\tablenotemark{a} & -0.20 & -0.19 $\pm$ 0.29 \\
        
        $M_1\sin{i_1}$ (\mearth) & 29.4 & 30.0 $\pm$ 6.2 \\
        \\
        \thisstar\,c & & \\
        $\sqrt{e_2}\cos{\omega_2}$ & -0.00 & -0.01 $\pm$ 0.09 \\
        $\sqrt{e_2}\sin{\omega_2}$ & -0.12 & -0.05 $\pm$ 0.04 \\
        
        $e_2$\tablenotemark{a} & 0.01 & 0.00 $\pm$ 0.01 \\
        
        $\omega_2$ (Radians)\tablenotemark{a} & -1.61 & -1.70 $\pm$ 1.46 \\
        
        $M_2\sin{i_2}$ (\mearth) & 191 & 195 $\pm$ 14 \\
       \\
        \thisstar\,d & & \\
        $\sqrt{e_3}\cos{\omega_3}$ & 0.44 & 0.44 $\pm$ 0.03 \\
        $\sqrt{e_3}\sin{\omega_3}$ & -0.12 & -0.12 $\pm$ 0.04 \\
        
        $e_3$\tablenotemark{a} & 0.21 & 0.20 $\pm$ 0.01 \\
        
        $\omega_3$ (Radians)\tablenotemark{a} & -0.27 & -0.26 $\pm$ 0.10 \\
        
        $M_3\sin{i_3}$ (\mearth) & 1767 & 1784 $\pm$ 120 \\
        $M_3\sin{i_3}$ (\mjup) & 5.55 & 5.61 $\pm$ 0.38 \\
        
        $P_3$ (days) & 1002 & 1002 $\pm$ 5 \\
        $t_{\textrm{tr},3}$ (BJD-2,400,000) & 56449 & 56450 $\pm$ 7\\ 
        \\
        System Parameters & & \\
        $\dot{\gamma}$ (m $s^{-1}$ $yr^{-1}$) & -0.26 & -0.25 $\pm$ 0.33 \\
        $\gamma_{\textrm{HARPS}}$ (m s$^{-1}$) & -54276.1 & -54276.2 $\pm$ 2.0 \\
        $\gamma_{\textrm{HIRES}}$ (m s$^{-1}$) & -27.7 & -27.7 $\pm$ 2.0 \\
        $\sigma_{jitter, \textrm{HARPS}}$ (m s$^{-1}$) & 0.72 & 1.23 $\pm$ 0.466 \\
        $\sigma_{jitter, \textrm{HIRES}}$ (m s$^{-1}$) & 1.68 & 1.80 $\pm$ 0.179 \\
        \tablenotetext{1}{Derived quantity}
    \end{tabular}
\end{table}

For \thisstar\,d itself, we return a 
Doppler semiamplitude of 95.21 $\pm$ 1.84 m s$^{-1}$, corresponding
to a minimum mass of $5.61 \pm 0.38$ \mjup\ ($1784 \pm 120$ \mearth).
We also measure a period of $1002 \pm 5$ days, an eccentricity of $0.20 \pm 0.01$, and a semimajor axis of $2.16 \pm 0.08$ AU.

\subsection{Limits on a Fourth Planet}
\label{sec:fourth}

A fourth planet beyond the orbit of 
\thisstar\,d, if it exists, could be observable 
through the detection of a long-term trend in the data. 
Given our three-year baseline of observations, we can place limits on the presence of such an outer companion.
From our \texttt{emcee} results, we find a long-term RV acceleration of $-0.25 \pm 0.32$ m s$^{-1}$ yr$^{-1}$. 
The $95^{th}$ percentile value of the \texttt{emcee} posterior probability distribution for $\dot{\gamma}$ provides an upper limit on acceleration from a fourth planet of 0.80 m s$^{-1}$ yr$^{-1}$.

From \citet{Montet14}, we know the maximum trend caused by a
planetary companion on a circular orbit is
\begin{equation}
\label{eq:trend}
\dot\gamma = (6.57 \textrm{m s$^{-1}$ yr$^{-1}$})\bigg(\frac{m_p \sin i}{M_J}\bigg)\bigg(\frac{a}{5\textrm{AU}}\bigg)^{-2},
\end{equation}
where $m_p$ is the mass of the planet, $M_J$ the mass of Jupiter, and $a$ the orbital semimajor axis.
From this, we can place limits on the presence of outer companions
with $m \sin i$ larger than $0.49 M_J$ at 10 AU and $1.95 M_J$ at 20 AU;
such companions must be at particular points in their orbits or at low inclination in order to evade RV detection.

At $V\simeq13$ mag, \thisstar\ falls just within \textit{Gaia}'s bright-star limit \citep{Perryman14}. A fourth planet's acceleration on \thisstar\ in \textit{Gaia} astrometry might be detectable at the level of 10-20 $\mu$as/yr$^2$ over the course of the mission. Averaging over flat priors for orbital angles and eccentricity, at the nominal distance of Kepler-56 ($d \sim 850$ pc), \textit{Gaia} could in principle detect curvature due to orbital motion of a companion of $\gtrsim20$ M$_\mathrm{J}$ at 10 AU or $\gtrsim80$ M$_\mathrm{J}$ at 20 AU. These values in Equation \ref{eq:trend} return, at the lowest, an acceleration of 32.85 m s$^{-1}$ yr$^{-1}$. This is much higher than the limits returned by our fit, suggesting that, save for face-on orbits, \textit{Gaia} will be less helpful than continued RV observation in placing further limits on a fourth planet.

A fourth planet in a near-resonant orbit with Kepler-56\,d could 
masquerade as a single eccentric planet, as described by \citet{Anglada-Escude10}.
However, we find the probability of this scenario to be low.
Re-running our \texttt{emcee} fit with the outer planet's eccentricity fixed at 0 leads to decreased likelihoods for the fit as a whole,
and we do not detect any long-term structure in the residuals.  However, our observations do not have the time resolution necessary to make a definitive assertion on this effect. Complete phase coverage of \thisstar\,d is needed to answer this question.

\section{Discussion}
\label{discussion}

\subsection{Comparison to Previous Work}
Our research supports that of \citet{Huber13} in finding strong evidence for a massive, non-transiting exoplanet in the Kepler-56 system. Now that our observations span a full \thisstar\,d orbit, we can compare our results with the projections from \citet{Huber13}, who predicted that both the planetary obliquity and long-term RV trend could both be broadly explained by a non-transiting companion with a period of $900$ days and mass $3.3$ \mjup.

Both our minimum mass and period are similar to the representative values listed by \citet{Huber13}. \thisstar\,d's minimum mass could be commensurate with that of a giant planet or a brown dwarf (for inclinations below 30 degrees). This could have implications for the near 2:1 resonance of the inner planets' orbits, as well as for the misalignment of their orbital plane with that of \thisstar\,'s rotation. Indeed, \citet{Li14} simulated several scenarios and found a higher probability of the observed misalignment being of a dynamical origin (e.g. \citealt{Fabrycky07}) than from migration of the bodies in a tilted protoplanetary disk (e.g. \citealt{Bate10}) or through angular momentum transport in the star itself that led to an apparent misalignment, even if the system was originally aligned (\citealt{Rogers12}).

While \thisstar\,d is a possible source of dynamical perturbation, Gratia \& Fabrycky (2016; submitted) simulate the scattering of two giant outer planets and find scattering between a system of three outer planets is required to excite the two inner planets of the system to inclinations similar to those observed in the data while preserving coplanarity.
These additional planets, if real, must be scattered to large orbital
separations or ejected entirely to evade detection by our RV observations.

\subsection{The Effect of Kepler-56 d on Transits of the Inner Planets}
\label{sec:transits}
\citet{Huber13} inferred masses of the system's inner planets by dynamically modeling their transits, ignoring possible
perturbations from the third, outer planet. We verify that this is a reasonable assumption by checking two effects that may be significant: a tidal term
corresponding to the change in the gravitational potential as 
\thisstar\,d completes its orbit, and a Roemer delay as the distance to the 
inner planets and host star vary over the orbit of the outer planet.

Following Equations 25-27 of \citet{Agol05},
the tidal effects would cause, over a long time baseline, the
transits of an inner planet with mass $m_1$ and period $P_1$ to be perturbed with a standard
deviation
\begin{equation}
\sigma = \frac{3\beta e_2}{\sqrt{2}(1-e_2^2)^{3/2}}\bigg[1-\frac{3}{16}e_2^2-\frac{47}{1296}e_2^4 - \frac{413}{27648}e^6_2\bigg]^{1/2},
\end{equation}
where $e_2$ is the eccentricity of the outer planet and 
\begin{equation}
\beta = \frac{m_2}{2\pi (m_0+m_1)} \frac{P_1^2}{P_2}.
\end{equation}
Here, $m_2$ is the mass of the outer planet with orbital period $P_2$, and $m_0$ the mass of the host star. 

For the values in Table \ref{tab:results} for our system, we find perturbations 
in the time of transit on the order of four seconds for \thisstar\,b and sixteen seconds for \thisstar\,c. Given that the 
precision in the measurement of times of transit of these planets
is typically tens of minutes, we do not expect these perturbations to affect, or be noticeable in, the measured times of transit.

The light travel time, or Roemer, delay is the result of changes in the physical distance between the observer and the host star
due to the orbit of the outer body. Following Equations 6 and 7 of \citet{Rappaport13}, its magnitude is bounded such that
\begin{equation}
A_R \leq \frac{G^{1/3}}{c(2\pi)^{2/3}}P_2^{2/3} \bigg[\frac{m_2 \sin i_2}{(m_0+m_1+m_2)^{2/3}}\bigg],
\end{equation}
where $G$ is Newton's constant, $c$ the speed of light, and all
other terms retain their meaning from the previous equation.
Inserting values from Table \ref{tab:results} again, we find an expected light travel time signal not to exceed 5 seconds, significantly smaller than the observed uncertainties, so we do not expect \thisstar\,d to affect the orbits of the inner planets in any observable way.

\subsection{Alternative Methods of Measuring Kepler-56\,d}

From our model, we measure a time of transit for \thisstar\,d of BJD-$2,400,000 = 56,450 \pm 7$ days.
Thus, if the planet transits the star, we would expect to detect a single
transit in the \kep\ dataset that is visible by eye, but
do not observe one in this window. 
As we know the posterior distribution of allowed
times of transit, we can determine the probability the planet
transited during a data gap.
In Quarters 6 and 7, there are four data gaps larger than 12 hours in which a transit
could reside.
Together, these gaps represent 1.7\% of the mass of the posterior distribution of the time of central transit.
The transit duration allows us to place even tighter constraints.
If \thisstar\,d transited with an impact parameter $b = 0$, the
transit would have a duration of 3.1 days. As none of the gaps are
longer than 20 hours in duration, we can additionally rule out any
transits with $b < 0.95$. By again integrating over the posterior
distribution but accounting for the nonzero transit duration, assuming
a flat distribution in impact parameter, we find that only 0.07\% of
allowed transits fall fully inside a data gap. If \thisstar\,d were
to transit, there is a 99.93\% proabability it would be observable in the \kep\ data.
Given this low probability and the \textit{a priori} small
transit probability for a companion on a $\sim 1000$ day period, it
is likely this companion is non-transiting.
We note that non-transiting does not necessarily imply non-coplanar
with the inner planets, as the transit probability decreases with
increasing semimajor axis \citep{Borucki84}.

Having measured the minimum mass ($m \sin i$) and orbital semimajor
axis ($a$) of \thisstar\,d, we can consider the possibility that the \textit{Gaia} astrometric mission would be able to constrain its inclination.
For lower (more face-on) inclinations, the planet will have a higher mass and
the center of mass of the system will move closer to the planet.
Additionally, the astrometric orbit will change shape on the sky, with more
face-on inclinations appearing more circular throughout an orbit.

Given the distance to the system ($d \sim 850$ pc) and the inferred semimajor
axis $2.13 \pm 0.07$ AU, the orbit of \thisstar\,d has a projected semimajor axis on the sky of $\sim 2.5$ mas. From the mass ratio between the planet and star, 
we then expect an astrometric signal with a semiamplitude of $11\sin^{-1} \mu$as.
\citet{Perryman14} determined that \textit{Gaia}
will detect planets with astrometric signatures larger than $68 \mu$as for stars as bright as \thisstar, meaning
this planet would evade detection at all except the lowest inclinations.
However, given that the \textit{Gaia} data can be combined with the prior information about the orbit of \thisstar\,d from RVs, it may be possible that
the planet will be detected at slightly lower inclinations. Regardless, the 
prospects of a robust determination of the outer planet's complete set of orbital parameters from 
\textit{Gaia} appear unlikely.

{\it Facilities:} \facility{Keck:I (HIRES)}, \facility{TNG (HARPS-N)}

\section{Summary}
\label{summary}

\thisstar, a red giant targeted in the telescope's primary mission, has a massive, non-transiting companion detected through radial velocities. This star is one of only a few red giants known to have transiting planets, and these planets orbit with a nearly 2:1 period ratio on a plane misaligned relative to the spin of their host star. The presence of another body in the system was first detected by \citet{Huber13} with observations from Keck/HIRES; we follow them up with subsequent observations from HIRES and HARPS-North at TNG. Incorporating these new data, we model the RV curve for a three-planet system. Our results confirm the existence of \thisstar\,d, with a period of $1002 \pm 5$ days and a minimum mass of $5.61 \pm 0.38$ \mjup. We also return an upper limit of acceleration from a possible fourth planet of 0.80 m s$^{-1}$ yr$^{-1}$ at 95\% confidence, severely restricting the possibility of the existence of other giant planets within $\sim 20$ AU.
We find that \thisstar\,d should not be detectable through its dynamical effect on the transits of the two inner planets, but for sufficiently face-on (more massive) orbits could be detectable through \textit{Gaia} observations of its astrometric wobble.

\acknowledgements

We thank Eric Agol, Daniel Fabrycky, and Daniel Huber for comments and conversations which improved the quality of this manuscript.

O.J.O. thanks the members and friends of the Banneker Institute, who made the summer in which this project began a fruitful time. He also thanks Neta Bahcall for allowing him to continue this research as his senior thesis. He gratefully acknowledges support from the Banneker Institute and Princeton's astrophysics department, Class of 1984, and Office of the Dean of Undergraduate Students in facilitating travel to AAS 227 to present this research. He would be remiss to forget the other members of the Party of Three and their associates.

B.T.M. is supported by the National Science Foundation Graduate Research Fellowship under Grant No. DGE‐1144469.

J.A.J. is supported by generous grants from the David and Lucile Packard Foundation and the Alfred P. Sloan Foundation.

C.A.W. acknowledges support from STFC grant ST/L000709/1.

This publication was made possible through the support of a grant from the John Templeton Foundation. The opinions expressed in this publication are those of the authors and do not necessarily reflect the views of the John Templeton Foundation. This material is based upon work supported by the National Aeronautics and Space Administration under grant No. NNX15AC90G issued through the Exoplanets Research Program.

The research leading to these results has received funding from the European Union Seventh Framework Programme (FP7/2007-2013) under Grant Agreement No. 313014 (ETAEARTH.)

Some of the data presented herein were obtained at the W.M. Keck Observatory, which is operated as a scientific partnership among the California Institute of Technology, the University of California and the National Aeronautics and Space Administration. The Observatory was made possible by the generous financial support of the W.M. Keck Foundation.
The authors wish to recognize and acknowledge the very significant cultural role and reverence that the summit of Maunakea has always had within the indigenous Hawaiian community.  We are most fortunate to have the opportunity to conduct observations from this mountain.

The HARPS-N project was funded by the Prodex program
of the Swiss Space Office (SSO), the Harvard University Origin of Life Initiative
(HUOLI), the Scottish Universities Physics Alliance (SUPA), the University
of Geneva, the Smithsonian Astrophysical Observatory (SAO), and the Italian
National Astrophysical Institute (INAF), University of St. Andrews, Queen’s
University Belfast, and University of Edinburgh.


\begin{thebibliography}{}
\expandafter\ifx\csname natexlab\endcsname\relax\def\natexlab#1{#1}\fi

\bibitem[{{Agol} {et~al.}(2005){Agol}, {Steffen}, {Sari}, \&
  {Clarkson}}]{Agol05}
{Agol}, E., {Steffen}, J., {Sari}, R., \& {Clarkson}, W. 2005, \mnras, 359, 567

\bibitem[{{Anglada-Escud{\'e}} {et~al.}(2010){Anglada-Escud{\'e}},
  {L{\'o}pez-Morales}, \& {Chambers}}]{Anglada-Escude10}
{Anglada-Escud{\'e}}, G., {L{\'o}pez-Morales}, M., \& {Chambers}, J.~E. 2010,
  \apj, 709, 168

\bibitem[{{Baranne} {et~al.}(1996){Baranne}, {Queloz}, {Mayor}, {Adrianzyk},
  {Knispel}, {Kohler}, {Lacroix}, {Meunier}, {Rimbaud}, \& {Vin}}]{Baranne96}
{Baranne}, A., {Queloz}, D., {Mayor}, M., {et~al.} 1996, \aaps, 119, 373

\bibitem[{{Barclay} {et~al.}(2015){Barclay}, {Quintana}, {Adams}, {Ciardi},
  {Huber}, {Foreman-Mackey}, {Montet}, \& {Caldwell}}]{Barclay15}
{Barclay}, T., {Quintana}, E.~V., {Adams}, F.~C., {et~al.} 2015, \apj, 809, 7

\bibitem[{{Batalha} {et~al.}(2010){Batalha}, {Borucki}, {Koch}, {Bryson},
  {Haas}, {Brown}, {Caldwell}, {Hall}, {Gilliland}, {Latham}, {Meibom}, \&
  {Monet}}]{Batalha10}
{Batalha}, N.~M., {Borucki}, W.~J., {Koch}, D.~G., {et~al.} 2010, \apjl, 713,
  L109

\bibitem[{{Batalha} {et~al.}(2013){Batalha}, {Rowe}, {Bryson}, {Barclay},
  {Burke}, {Caldwell}, {Christiansen}, {Mullally}, {Thompson}, {Brown},
  {Dupree}, {Fabrycky}, {Ford}, {Fortney}, {Gilliland}, {Isaacson}, {Latham},
  {Marcy}, {Quinn}, {Ragozzine}, {Shporer}, {Borucki}, {Ciardi}, {Gautier},
  {Haas}, {Jenkins}, {Koch}, {Lissauer}, {Rapin}, {Basri}, {Boss}, {Buchhave},
  {Carter}, {Charbonneau}, {Christensen-Dalsgaard}, {Clarke}, {Cochran},
  {Demory}, {Desert}, {Devore}, {Doyle}, {Esquerdo}, {Everett}, {Fressin},
  {Geary}, {Girouard}, {Gould}, {Hall}, {Holman}, {Howard}, {Howell},
  {Ibrahim}, {Kinemuchi}, {Kjeldsen}, {Klaus}, {Li}, {Lucas}, {Meibom},
  {Morris}, {Pr{\v s}a}, {Quintana}, {Sanderfer}, {Sasselov}, {Seader},
  {Smith}, {Steffen}, {Still}, {Stumpe}, {Tarter}, {Tenenbaum}, {Torres},
  {Twicken}, {Uddin}, {Van Cleve}, {Walkowicz}, \& {Welsh}}]{Batalha13}
{Batalha}, N.~M., {Rowe}, J.~F., {Bryson}, S.~T., {et~al.} 2013, \apjs, 204, 24

\bibitem[{{Bate} {et~al.}(2010){Bate}, {Lodato}, \& {Pringle}}]{Bate10}
{Bate}, M.~R., {Lodato}, G., \& {Pringle}, J.~E. 2010, \mnras, 401, 1505

\bibitem[{{Borucki} \& {Summers}(1984)}]{Borucki84}
{Borucki}, W.~J., \& {Summers}, A.~L. 1984, Icarus, 58, 121

\bibitem[{{Borucki} {et~al.}(2010){Borucki}, {Koch}, {Basri}, {Batalha},
  {Brown}, {Caldwell}, {Caldwell}, {Christensen-Dalsgaard}, {Cochran},
  {DeVore}, {Dunham}, {Dupree}, {Gautier}, {Geary}, {Gilliland}, {Gould},
  {Howell}, {Jenkins}, {Kondo}, {Latham}, {Marcy}, {Meibom}, {Kjeldsen},
  {Lissauer}, {Monet}, {Morrison}, {Sasselov}, {Tarter}, {Boss}, {Brownlee},
  {Owen}, {Buzasi}, {Charbonneau}, {Doyle}, {Fortney}, {Ford}, {Holman},
  {Seager}, {Steffen}, {Welsh}, {Rowe}, {Anderson}, {Buchhave}, {Ciardi},
  {Walkowicz}, {Sherry}, {Horch}, {Isaacson}, {Everett}, {Fischer}, {Torres},
  {Johnson}, {Endl}, {MacQueen}, {Bryson}, {Dotson}, {Haas}, {Kolodziejczak},
  {Van Cleve}, {Chandrasekaran}, {Twicken}, {Quintana}, {Clarke}, {Allen},
  {Li}, {Wu}, {Tenenbaum}, {Verner}, {Bruhweiler}, {Barnes}, \&
  {Prsa}}]{Borucki10}
{Borucki}, W.~J., {Koch}, D., {Basri}, G., {et~al.} 2010, Science, 327, 977

\bibitem[{{Borucki} {et~al.}(2011){Borucki}, {Koch}, {Basri}, {Batalha},
  {Brown}, {Bryson}, {Caldwell}, {Christensen-Dalsgaard}, {Cochran}, {DeVore},
  {Dunham}, {Gautier}, {Geary}, {Gilliland}, {Gould}, {Howell}, {Jenkins},
  {Latham}, {Lissauer}, {Marcy}, {Rowe}, {Sasselov}, {Boss}, {Charbonneau},
  {Ciardi}, {Doyle}, {Dupree}, {Ford}, {Fortney}, {Holman}, {Seager},
  {Steffen}, {Tarter}, {Welsh}, {Allen}, {Buchhave}, {Christiansen}, {Clarke},
  {Das}, {D{\'e}sert}, {Endl}, {Fabrycky}, {Fressin}, {Haas}, {Horch},
  {Howard}, {Isaacson}, {Kjeldsen}, {Kolodziejczak}, {Kulesa}, {Li}, {Lucas},
  {Machalek}, {McCarthy}, {MacQueen}, {Meibom}, {Miquel}, {Prsa}, {Quinn},
  {Quintana}, {Ragozzine}, {Sherry}, {Shporer}, {Tenenbaum}, {Torres},
  {Twicken}, {Van Cleve}, {Walkowicz}, {Witteborn}, \& {Still}}]{Borucki11b}
{Borucki}, W.~J., {Koch}, D.~G., {Basri}, G., {et~al.} 2011, \apj, 736, 19

\bibitem[{{Burke} {et~al.}(2014){Burke}, {Bryson}, {Mullally}, {Rowe},
  {Christiansen}, {Thompson}, {Coughlin}, {Haas}, {Batalha}, {Caldwell},
  {Jenkins}, {Still}, {Barclay}, {Borucki}, {Chaplin}, {Ciardi}, {Clarke},
  {Cochran}, {Demory}, {Esquerdo}, {Gautier}, {Gilliland}, {Girouard}, {Havel},
  {Henze}, {Howell}, {Huber}, {Latham}, {Li}, {Morehead}, {Morton}, {Pepper},
  {Quintana}, {Ragozzine}, {Seader}, {Shah}, {Shporer}, {Tenenbaum}, {Twicken},
  \& {Wolfgang}}]{Burke14}
{Burke}, C.~J., {Bryson}, S.~T., {Mullally}, F., {et~al.} 2014, \apjs, 210, 19

\bibitem[{{Butler} {et~al.}(1996){Butler}, {Marcy}, {Williams}, {McCarthy},
  {Dosanjh}, \& {Vogt}}]{Butler_etal96}
{Butler}, R.~P., {Marcy}, G.~W., {Williams}, E., {et~al.} 1996, \pasp, 108, 500

\bibitem[{{Ciceri} {et~al.}(2015){Ciceri}, {Lillo-Box}, {Southworth},
  {Mancini}, {Henning}, \& {Barrado}}]{Ciceri15}
{Ciceri}, S., {Lillo-Box}, J., {Southworth}, J., {et~al.} 2015, \aap, 573, L5

\bibitem[{{Cosentino} {et~al.}(2012){Cosentino}, {Lovis}, {Pepe}, {Collier
  Cameron}, {Latham}, {Molinari}, {Udry}, {Bezawada}, {Black}, {Born},
  {Buchschacher}, {Charbonneau}, {Figueira}, {Fleury}, {Galli}, {Gallie},
  {Gao}, {Ghedina}, {Gonzalez}, {Gonzalez}, {Guerra}, {Henry}, {Horne},
  {Hughes}, {Kelly}, {Lodi}, {Lunney}, {Maire}, {Mayor}, {Micela}, {Ordway},
  {Peacock}, {Phillips}, {Piotto}, {Pollacco}, {Queloz}, {Rice}, {Riverol},
  {Riverol}, {San Juan}, {Sasselov}, {Segransan}, {Sozzetti}, {Sosnowska},
  {Stobie}, {Szentgyorgyi}, {Vick}, \& {Weber}}]{Cosentino12}
{Cosentino}, R., {Lovis}, C., {Pepe}, F., {et~al.} 2012, in \procspie, Vol.
  8446, Ground-based and Airborne Instrumentation for Astronomy IV, 84461V

\bibitem[{{Eastman} {et~al.}(2013){Eastman}, {Gaudi}, \& {Agol}}]{Eastman13}
{Eastman}, J., {Gaudi}, B.~S., \& {Agol}, E. 2013, \pasp, 125, 83

\bibitem[{{Fabrycky} \& {Tremaine}(2007)}]{Fabrycky07}
{Fabrycky}, D., \& {Tremaine}, S. 2007, \apj, 669, 1298

\bibitem[{{Ford} {et~al.}(2011){Ford}, {Rowe}, {Fabrycky}, {Carter}, {Holman},
  {Lissauer}, {Ragozzine}, {Steffen}, {Batalha}, {Borucki}, {Bryson},
  {Caldwell}, {Dunham}, {Gautier}, {Jenkins}, {Koch}, {Li}, {Lucas}, {Marcy},
  {McCauliff}, {Mullally}, {Quintana}, {Still}, {Tenenbaum}, {Thompson}, \&
  {Twicken}}]{Ford11}
{Ford}, E.~B., {Rowe}, J.~F., {Fabrycky}, D.~C., {et~al.} 2011, \apjs, 197, 2

\bibitem[{{Ford} {et~al.}(2012){Ford}, {Fabrycky}, {Steffen}, {Carter},
  {Fressin}, {Holman}, {Lissauer}, {Moorhead}, {Morehead}, {Ragozzine}, {Rowe},
  {Welsh}, {Allen}, {Batalha}, {Borucki}, {Bryson}, {Buchhave}, {Burke},
  {Caldwell}, {Charbonneau}, {Clarke}, {Cochran}, {D{\'e}sert}, {Endl},
  {Everett}, {Fischer}, {Gautier}, {Gilliland}, {Jenkins}, {Haas}, {Horch},
  {Howell}, {Ibrahim}, {Isaacson}, {Koch}, {Latham}, {Li}, {Lucas}, {MacQueen},
  {Marcy}, {McCauliff}, {Mullally}, {Quinn}, {Quintana}, {Shporer}, {Still},
  {Tenenbaum}, {Thompson}, {Torres}, {Twicken}, {Wohler}, \& {the Kepler
  Science Team}}]{Ford12TTV}
{Ford}, E.~B., {Fabrycky}, D.~C., {Steffen}, J.~H., {et~al.} 2012, \apj, 750,
  113

\bibitem[{{Foreman-Mackey} {et~al.}(2013){Foreman-Mackey}, {Hogg}, {Lang}, \&
  {Goodman}}]{Foreman-Mackey13}
{Foreman-Mackey}, D., {Hogg}, D.~W., {Lang}, D., \& {Goodman}, J. 2013, \pasp,
  125, 306

\bibitem[{Goodman \& Weare(2010)}]{Goodman10}
Goodman, J., \& Weare, J. 2010, Communications in Applied Mathematics and
  Computational Science, 5, 65

\bibitem[{{Grunblatt} {et~al.}(2016){Grunblatt}, {Huber}, {Gaidos}, {Lopez},
  {Fulton}, {Fortney}, {Howard}, {Sinukoff}, {Mann}, \&
  {Isaacson}}]{Grunblatt16}
{Grunblatt}, S.~K., {Huber}, D., {Gaidos}, E.~J., {et~al.} 2016, ArXiv
  e-prints, arXiv:1606.05818

\bibitem[{{Howard} {et~al.}(2010){Howard}, {Johnson}, {Marcy}, {Fischer},
  {Wright}, {Bernat}, {Henry}, {Peek}, {Isaacson}, {Apps}, {Endl}, {Cochran},
  {Valenti}, {Anderson}, \& {Piskunov}}]{Howard10}
{Howard}, A.~W., {Johnson}, J.~A., {Marcy}, G.~W., {et~al.} 2010, \apj, 721,
  1467

\bibitem[{{Huber} {et~al.}(2013){Huber}, {Carter}, {Barbieri}, {Miglio},
  {Deck}, {Fabrycky}, {Montet}, {Buchhave}, {Chaplin}, {Hekker},
  {Montalb{\'a}n}, {Sanchis-Ojeda}, {Basu}, {Bedding}, {Campante},
  {Christensen-Dalsgaard}, {Elsworth}, {Stello}, {Arentoft}, {Ford},
  {Gilliland}, {Handberg}, {Howard}, {Isaacson}, {Johnson}, {Karoff},
  {Kawaler}, {Kjeldsen}, {Latham}, {Lund}, {Lundkvist}, {Marcy}, {Metcalfe},
  {Silva Aguirre}, \& {Winn}}]{Huber13}
{Huber}, D., {Carter}, J.~A., {Barbieri}, M., {et~al.} 2013, Science, 342, 331

\bibitem[{{Johnson}(2008)}]{Johnson08}
{Johnson}, J.~A. 2008, in Astronomical Society of the Pacific Conference
  Series, Vol. 398, Extreme Solar Systems, ed. D.~{Fischer}, F.~A. {Rasio},
  S.~E. {Thorsett}, \& A.~{Wolszczan}, 59

\bibitem[{{Kipping} {et~al.}(2016){Kipping}, {Torres}, {Henze}, {Teachey},
  {Isaacson}, {Petigura}, {Marcy}, {Buchhave}, {Chen}, {Bryson}, \&
  {Sandford}}]{Kipping16}
{Kipping}, D.~M., {Torres}, G., {Henze}, C., {et~al.} 2016, \apj, 820, 112

\bibitem[{{Kostov} {et~al.}(2015){Kostov}, {Orosz}, {Welsh}, {Doyle},
  {Fabrycky}, {Haghighipour}, {Quarles}, {Short}, {Cochran}, {Endl}, {Ford},
  {Gregorio}, {Hinse}, {Isaacson}, {Jenkins}, {Jensen}, {Kane}, {Kull},
  {Latham}, {Lissauer}, {Marcy}, {Mazeh}, {Muller}, {Pepper}, {Quinn},
  {Ragozzine}, {Shporer}, {Steffen}, {Torres}, {Windmiller}, \&
  {Borucki}}]{Kostov16}
{Kostov}, V.~B., {Orosz}, J.~A., {Welsh}, W.~F., {et~al.} 2015, ArXiv e-prints,
  arXiv:1512.00189

\bibitem[{{Lehmann-Filh{\'e}s}(1894)}]{LehmannFilhes94}
{Lehmann-Filh{\'e}s}, R. 1894, Astronomische Nachrichten, 136, 17

\bibitem[{{Li} {et~al.}(2014){Li}, {Naoz}, {Valsecchi}, {Johnson}, \&
  {Rasio}}]{Li14}
{Li}, G., {Naoz}, S., {Valsecchi}, F., {Johnson}, J.~A., \& {Rasio}, F.~A.
  2014, \apj, 794, 131

\bibitem[{{Lillo-Box} {et~al.}(2014){Lillo-Box}, {Barrado}, {Moya},
  {Montesinos}, {Montalb{\'a}n}, {Bayo}, {Barbieri}, {R{\'e}gulo}, {Mancini},
  {Bouy}, \& {Henning}}]{Lillo-Box14}
{Lillo-Box}, J., {Barrado}, D., {Moya}, A., {et~al.} 2014, \aap, 562, A109

\bibitem[{{Montet} {et~al.}(2014){Montet}, {Crepp}, {Johnson}, {Howard}, \&
  {Marcy}}]{Montet14}
{Montet}, B.~T., {Crepp}, J.~R., {Johnson}, J.~A., {Howard}, A.~W., \& {Marcy},
  G.~W. 2014, \apj, 781, 28

\bibitem[{{Mullally} {et~al.}(2015){Mullally}, {Coughlin}, {Thompson}, {Rowe},
  {Burke}, {Latham}, {Batalha}, {Bryson}, {Christiansen}, {Henze}, {Ofir},
  {Quarles}, {Shporer}, {Van Eylen}, {Van Laerhoven}, {Shah}, {Wolfgang},
  {Chaplin}, {Xie}, {Akeson}, {Argabright}, {Bachtell}, {Barclay}, {Borucki},
  {Caldwell}, {Campbell}, {Catanzarite}, {Cochran}, {Duren}, {Fleming},
  {Fraquelli}, {Girouard}, {Haas}, {He{\l}miniak}, {Howell}, {Huber}, {Larson},
  {Gautier}, {Jenkins}, {Li}, {Lissauer}, {McArthur}, {Miller}, {Morris},
  {Patil-Sabale}, {Plavchan}, {Putnam}, {Quintana}, {Ramirez}, {Silva Aguirre},
  {Seader}, {Smith}, {Steffen}, {Stewart}, {Stober}, {Still}, {Tenenbaum},
  {Troeltzsch}, {Twicken}, \& {Zamudio}}]{Mullaly15}
{Mullally}, F., {Coughlin}, J.~L., {Thompson}, S.~E., {et~al.} 2015, \apjs,
  217, 31

\bibitem[{{Pepe} {et~al.}(2002){Pepe}, {Mayor}, {Rupprecht}, {Avila},
  {Ballester}, {Beckers}, {Benz}, {Bertaux}, {Bouchy}, {Buzzoni}, {Cavadore},
  {Deiries}, {Dekker}, {Delabre}, {D'Odorico}, {Eckert}, {Fischer}, {Fleury},
  {George}, {Gilliotte}, {Gojak}, {Guzman}, {Koch}, {Kohler}, {Kotzlowski},
  {Lacroix}, {Le Merrer}, {Lizon}, {Lo Curto}, {Longinotti}, {Megevand},
  {Pasquini}, {Petitpas}, {Pichard}, {Queloz}, {Reyes}, {Richaud}, {Sivan},
  {Sosnowska}, {Soto}, {Udry}, {Ureta}, {van Kesteren}, {Weber}, {Weilenmann},
  {Wicenec}, {Wieland}, {Christensen-Dalsgaard}, {Dravins}, {Hatzes},
  {K{\"u}rster}, {Paresce}, \& {Penny}}]{Pepe02}
{Pepe}, F., {Mayor}, M., {Rupprecht}, G., {et~al.} 2002, The Messenger, 110, 9

\bibitem[{{Pepper} {et~al.}(2016){Pepper}, {Rodriguez}, {Collins}, {Johnson},
  {Fulton}, {Howard}, {Beatty}, {Stassun}, {Isaacson}, {Col{\'o}n}, {Lund},
  {Kuhn}, {Siverd}, {Gaudi}, {Tan}, {Curtis}, {Stockdale}, {Mawet}, {Bottom},
  {James}, {Zhou}, {Bayliss}, {Cargile}, {Bieryla}, {Penev}, {Latham},
  {Labadie-Bartz}, {Kielkopf}, {Eastman}, {Oberst}, {Jensen}, {Nelson},
  {Sliski}, {Wittenmyer}, {McCrady}, {Wright}, \& {Relles}}]{Pepper16}
{Pepper}, J., {Rodriguez}, J.~E., {Collins}, K.~A., {et~al.} 2016, ArXiv
  e-prints, arXiv:1607.01755

\bibitem[{{Perryman} {et~al.}(2014){Perryman}, {Hartman}, {Bakos}, \&
  {Lindegren}}]{Perryman14}
{Perryman}, M., {Hartman}, J., {Bakos}, G.~{\'A}., \& {Lindegren}, L. 2014,
  \apj, 797, 14

\bibitem[{{Quinn} {et~al.}(2015){Quinn}, {White}, {Latham}, {Chaplin},
  {Handberg}, {Huber}, {Kipping}, {Payne}, {Jiang}, {Silva Aguirre}, {Stello},
  {Sliski}, {Ciardi}, {Buchhave}, {Bedding}, {Davies}, {Hekker}, {Kjeldsen},
  {Kuszlewicz}, {Everett}, {Howell}, {Basu}, {Campante},
  {Christensen-Dalsgaard}, {Elsworth}, {Karoff}, {Kawaler}, {Lund},
  {Lundkvist}, {Esquerdo}, {Calkins}, \& {Berlind}}]{Quinn15}
{Quinn}, S.~N., {White}, T.~R., {Latham}, D.~W., {et~al.} 2015, \apj, 803, 49

\bibitem[{{Rappaport} {et~al.}(2013){Rappaport}, {Deck}, {Levine}, {Borkovits},
  {Carter}, {El Mellah}, {Sanchis-Ojeda}, \& {Kalomeni}}]{Rappaport13}
{Rappaport}, S., {Deck}, K., {Levine}, A., {et~al.} 2013, \apj, 768, 33

\bibitem[{{Rogers} {et~al.}(2012){Rogers}, {Lin}, \& {Lau}}]{Rogers12}
{Rogers}, T.~M., {Lin}, D.~N.~C., \& {Lau}, H.~H.~B. 2012, \apjl, 758, L6

\bibitem[{{Rowe} {et~al.}(2015){Rowe}, {Coughlin}, {Antoci}, {Barclay},
  {Batalha}, {Borucki}, {Burke}, {Bryson}, {Caldwell}, {Campbell},
  {Catanzarite}, {Christiansen}, {Cochran}, {Gilliland}, {Girouard}, {Haas},
  {He{\l}miniak}, {Henze}, {Hoffman}, {Howell}, {Huber}, {Hunter},
  {Jang-Condell}, {Jenkins}, {Klaus}, {Latham}, {Li}, {Lissauer}, {McCauliff},
  {Morris}, {Mullally}, {Ofir}, {Quarles}, {Quintana}, {Sabale}, {Seader},
  {Shporer}, {Smith}, {Steffen}, {Still}, {Tenenbaum}, {Thompson}, {Twicken},
  {Van Laerhoven}, {Wolfgang}, \& {Zamudio}}]{Rowe15}
{Rowe}, J.~F., {Coughlin}, J.~L., {Antoci}, V., {et~al.} 2015, \apjs, 217, 16

\bibitem[{{Steffen} {et~al.}(2012){Steffen}, {Ford}, {Rowe}, {Fabrycky},
  {Holman}, {Welsh}, {Batalha}, {Borucki}, {Bryson}, {Caldwell}, {Ciardi},
  {Jenkins}, {Kjeldsen}, {Koch}, {Pr{\v s}a}, {Sanderfer}, {Seader}, \&
  {Twicken}}]{Steffen12}
{Steffen}, J.~H., {Ford}, E.~B., {Rowe}, J.~F., {et~al.} 2012, \apj, 756, 186

\bibitem[{{Steffen} {et~al.}(2013){Steffen}, {Fabrycky}, {Agol}, {Ford},
  {Morehead}, {Cochran}, {Lissauer}, {Adams}, {Borucki}, {Bryson}, {Caldwell},
  {Dupree}, {Jenkins}, {Robertson}, {Rowe}, {Seader}, {Thompson}, \&
  {Twicken}}]{Steffen13}
{Steffen}, J.~H., {Fabrycky}, D.~C., {Agol}, E., {et~al.} 2013, \mnras, 428,
  1077

\bibitem[{{Valenti} {et~al.}(1995){Valenti}, {Butler}, \& {Marcy}}]{Valenti95}
{Valenti}, J.~A., {Butler}, R.~P., \& {Marcy}, G.~W. 1995, \pasp, 107, 966

\bibitem[{{Van Eylen} {et~al.}(2016){Van Eylen}, {Albrecht}, {Gandolfi}, {Dai},
  {Winn}, {Hirano}, {Narita}, {Bruntt}, {Prieto-Arranz}, {Bejar}, {Nowak},
  {Lund}, {Palle}, {Ribas}, {Sanchis-Ojeda}, {Yu}, {Arriagada}, {Butler},
  {Crane}, {Handberg}, {Deeg}, {Jessen-Hansen}, {Johnson}, {Nespral}, {Rogers},
  {Ryu}, {Shectman}, {Shrotriya}, {Slumstrup}, {Takeda}, {Teske}, {Thompson},
  {Vanderburg}, \& {Wittenmyer}}]{VanEylen16}
{Van Eylen}, V., {Albrecht}, S., {Gandolfi}, D., {et~al.} 2016, ArXiv e-prints,
  arXiv:1605.09180

\bibitem[{{Vogt} {et~al.}(1994){Vogt}, {Allen}, {Bigelow}, {Bresee}, {Brown},
  {Cantrall}, {Conrad}, {Couture}, {Delaney}, {Epps}, {Hilyard}, {Hilyard},
  {Horn}, {Jern}, {Kanto}, {Keane}, {Kibrick}, {Lewis}, {Osborne},
  {Pardeilhan}, {Pfister}, {Ricketts}, {Robinson}, {Stover}, {Tucker}, {Ward},
  \& {Wei}}]{Vogt94}
{Vogt}, S.~S., {Allen}, S.~L., {Bigelow}, B.~C., {et~al.} 1994, in Society of
  Photo-Optical Instrumentation Engineers (SPIE) Conference Series, Vol. 2198,
  Society of Photo-Optical Instrumentation Engineers (SPIE) Conference Series,
  ed. D.~L. {Crawford} \& E.~R. {Craine}, 362

\end{thebibliography}
\end{document}